\documentclass[twocolumn, noshowpacs, nopreprintnumbers, amsmath,amssymb,superscriptaddress]{revtex4}

\usepackage[svgnames]{xcolor}
\usepackage{hyperref}
\hypersetup{
  colorlinks = true,
  linkcolor =red,
  anchorcolor = red,
  citecolor = blue,
  urlcolor = blue
}
\usepackage{graphicx}
\usepackage{dcolumn}
\usepackage{bm}

\usepackage{appendix}
 \usepackage{verbatim}
\newtheorem{remark}{Remark}

\usepackage{breakurl}
\usepackage{amsmath}
\usepackage{enumerate}
\usepackage[addedmarkup=uline,bf]{changes}
\usepackage{epstopdf}
\usepackage{changes} %
\colorlet{Changes@Color}{red}  

\makeatletter
  \let\@font@info\@gobble
  \let\@font@warning\@gobble
\makeatother

\begin{document}
\hyphenation{maxi-mum}
\preprint{APS/123-QED}

\title{Transport in disordered systems: the single big jump approach}



\author{Wanli Wang}%
\affiliation{%
Department of Physics, Bar-Ilan University, Ramat-Gan
52900, Israel }
\affiliation{%
Institute of Nanotechnology and Advanced Materials, Bar-Ilan University, Ramat-Gan
52900, Israel }
\author{Alessandro Vezzani}%
\author{Raffaella Burioni}%
\affiliation{
 INFN, Gruppo Collegato di Parma, via G.P. Usberti 7/A-43124 Parma, Italy}
\author{Eli Barkai}%
\affiliation{%
Department of Physics, Bar-Ilan University, Ramat-Gan
52900, Israel }
\affiliation{%
Institute of Nanotechnology and Advanced Materials, Bar-Ilan University, Ramat-Gan
52900, Israel }


\date{\today}
\begin{abstract}
{In a growing number of strongly disordered and dense systems, the dynamics of a particle pulled by an external force field exhibits super-diffusion. In the context of glass forming systems, super cooled  glasses and contamination spreading in porous medium it was suggested to model this behavior with a biased continuous time random walk. Here we analyze the plume of particles far lagging behind the mean, with the single big jump principle. Revealing the mechanism  of the anomaly, we show how a single trapping time, the largest one, is responsible for the rare fluctuations in the system.  These non typical fluctuations still  control the behavior of the mean square displacement, which is the most basic quantifier of the dynamics  in many experimental setups.
We show how the initial conditions, describing either stationary state or non-equilibrium case, persist for ever in the sense that the rare fluctuations are sensitive to the initial preparation. To describe the fluctuations of the largest trapping time, we modify Fr\'{e}chet's law from extreme value statistics, taking into consideration the fact that the large fluctuations are
very different  from those observed for independent and identically distributed random variables. }
\end{abstract}

%

%


\maketitle
\section{Introduction}
\begin{figure}[h]
  \centering
  \includegraphics[width=9cm]{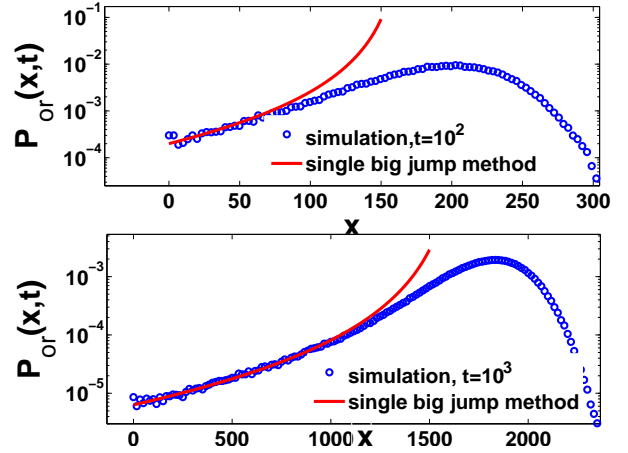}\\
  \caption{ The density of positions of particles for an ordinary CTRW model.  The spreading packet is non-Gaussian.
 The left plume of particles is due to the long  trapping times, which implies  that some particles are moving by far slower if compared with  the mean $\langle x(t)\rangle$. Somewhat paradoxically, these slow particles lead to super-diffusion as the MSD grows like $t^{3-\alpha}$ \cite{Shlesinger1974Asymptotic}.
  In this work we show how rare events  in this process are determined by the largest trapping times. In turn, it controls the behavior of the MSD. The typical fluctuations  are defined for $x\sim \langle x(t)\rangle$, i.e., close to the peak of the packet, while we focus on the rare fluctuations shown by the red solid line. The parameters are $a=5$, $\sigma=1$, and $\alpha=1.5$; see Eqs. \eqref{18eq104} and \eqref{phit}.
}
\label{trajectory}
\end{figure}

Diffusion and transport in a vast number of weakly disordered systems follows Gaussian statistics. As a consequence, the packet of the spreading particles  is symmetrically spread with respect to (w.r.t.) the mean $\langle x(t)\rangle$. In contrast, for strongly disordered systems, the packet  is found to be non-Gaussian and non-symmetric \cite{Shlesinger1974Asymptotic,Gradenigo2016Field}. Starting on $x=0$, the slowest particles are  trapped by the disorder, resulting in a plume of particles far lagging behind the mean $\langle x(t)\rangle$, i.e., the fluctuations are large and break symmetry (see Fig.~\ref{trajectory}).  Deep energetic and entropic traps, which hinder the motion are expected to lead to a slow down of the diffusion. The most frequently used quantifier of  diffusion processes is clearly the mean square displacement (MSD). However, in the presence of deep traps, the MSD exhibits super-diffusion. This is not an indication for a fast process, instead it is due to the very slow particles far lagging behind the mean, which lead to very large fluctuations of displacements.
Thus slow dynamics of a minority of particles leads to enhanced fluctuations and symmetry breaking  w.r.t. $\langle x(t)\rangle$. Such processes are widespread, in particular many works focused on the surprising discovery of the super-diffusion in dense environments \cite{Lechenault2010Super,Winter2012Active,Benichou2013Geometry,Schroer2013Anomalous,Leitmann2017Time,Pierre2018Nonequilibrium}. This was originally investigated in the context of diffusion in disordered material \cite{Shlesinger1974Asymptotic,Bouchaud1990Classical,Bouchaud1990Anomalous,Lechenault2010Super,Gradenigo2016Field,Akimoto2018Ergodicity,Hou2018Biased}, contamination  spreading in porous medium \cite{Brian1997Anomalous,Berkowitz2006Modeling,Yong2016Backward,Alon2017Time}, simulation of biased particles in glass forming systems  \cite{Winter2012Active}
and super cooled liquids \cite{Schroer2013Anomalous}, pulled by a constant force.

 \begin{figure}[h]
  \centering
  \includegraphics[width=9cm]{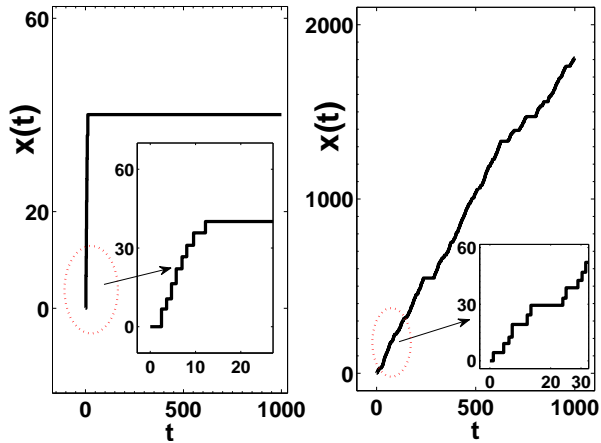}\\
  \caption{ Two trajectories of particles  ending at small and large $x$ when $t=1000$.
For the case where $x(t=1000)$ is near the original position, we see a very long waiting time, as the particle is trapped for a time of the order of $t$. In contrast, when $x(t=1000)\simeq \langle x(t)\rangle$, the trapping times are relatively short and comparable with each other. The inset shows the trajectory of the particle at a short time.
%
 The parameters are the same as in Fig.~\ref{trajectory}.
}
\label{trajectoryv2}
\end{figure}

Here we investigate the spreading of the packet of particles, using the biased continuous time random walk (CTRW) \cite{Metzler2000random,Klafter2011First,Kutner2017continuous}. Our goal is to characterize
precisely the mechanism leading to the large fluctuations. We promote the idea of the single big jump principle. This means that one and only one trapping time is responsible for the rare fluctuations.
Thus in this work we show the relation between the theory of extreme value statistics and the anomalous transport. For that we need to modify
the well-known Fr{\'e}chet law \cite{Gumbel2004Statistics,Albeverio2006Extreme} which describes extreme
events for uncorrelated systems. Similarly we present an analysis of the far tail of the spreading of the packet of particles, showing the deviations from the L{\'e}vy statistics describing the
bulk statistics. This is done for both non-stationary and equilibrium initial conditions. While the typical fluctuations in our systems are not sensitive to the initial conditions, the rare fluctuations do, and this we believe is a general theme for systems with fat-tailed statistics.

We will relate the position of the random walker $x(t)$ and the longest trapping interval $\tau_{\max}$. The typical fluctuations of both observables were considered previously, and were shown to behave as if they are composed of independent and identically distributed (IID) events, namely the L\'{e}vy stable law and Fr\'{e}chet's law hold for typical fluctuations (Eqs.~\eqref{19seq3} and \eqref{19seq7} below). We show below how these laws must be corrected when dealing with the far tail.
In turn standard Cramer's theorem from large deviation theory \cite{Touchette2009large}, which identifies the large fluctuations with the accumulation of many small steps, fails in this case studied here. More precisely, we claim below that one can obtain two limiting laws both for $x(t)$ and $\tau_{\max}$, the first is the just mentioned L\'{e}vy, Fr\'{e}chet laws and the second is an infinite density, i.e., a non-normalized state.

What is the principle of big jump? Many works have focus on the dominance of one big jump in a stochastic  process. For example consider the activation process of a particle over a barrier, modeled with an over-damped Langevin equation. If the noise is non-correlated and Gaussian, this escape is achieved by many small displacements, accumulating to give the rare escape from the well. On the other hand, if the noise is of the L\'{e}vy type, one event giving rise to a large fluctuation dominates the escape \cite{Karol2019Peculiarities}. Similar ideas hold for the analysis of random partition functions and were used in the study of the Sinai model \cite{Oshanin1993Behavior,Gleb2013Anomaous}.  In the context of a run-and-tumble model and combination phenomenon these insights are well understood \cite{Giacomo2017Participation,Gradenigo2019first}. Roughly speaking, one can see that the largest summand is of the order of the total sum, a theme which is already known.

To  be more  specific consider $N$ random variables $\{\vartheta_1,\vartheta_2,\cdots,\vartheta_N\}$. Let $\vartheta_{\max}$ be the maximum of the set and $S_{N}=\sum_{i=1}^N\vartheta_i$ is the sum. The dominance effect, found for example if $\vartheta_i$ are IID random variables drawn from a fat-tailed distribution, is the claim that $S_N$ and $\vartheta_{\max}$ are of the same order \cite{Cistjakov1964theorem}. More exactly, $S_N$ and $\vartheta_{\max}$ scale with $N$ the same way.
A more profound case is when the distribution of $\vartheta_{\max}$ is the same as that of $S_N$, besides a trivial constant and in a limit to be specified later. This is what we and others refer to as the principle of big jump. This statement was shown to be valid for sub-exponential  IID variables \cite{Cistjakov1964theorem} and see also \cite{Thomas2013Precise,Buraczewski2013Large}.  In the IID case, the statement is valid for any $N$, so the limit $N\to\infty$ is not at all required. Here our aim is show how the big jump principle holds for diffusion in disorder systems using the CTRW model. We will modify the principle to discuss the largest trapping time and its relation to the position of the random walker, so the principle discussed below is very different if compared to the original, in particular we depart form the IID case.


In \cite{Alessandro2019Single,Vezzani2019Microscopic}, we promoted a rate method to the big jump approach which was used to predict non-analytical behaviors of the far tail of L\'{e}vy walk process and the so called quenched L\'{e}vy-Lonentz gas model. In these works, the very basic approach is different from what we have here; see Eq.~\eqref{19seq2} below. Further the connection to the modified Fr\'{e}chet law, and the difference between stationary and non-equilibrium initial conditions are discussed here for the first time.

%
%

 \begin{figure}[htb]
  \centering
  \includegraphics[width=9cm]{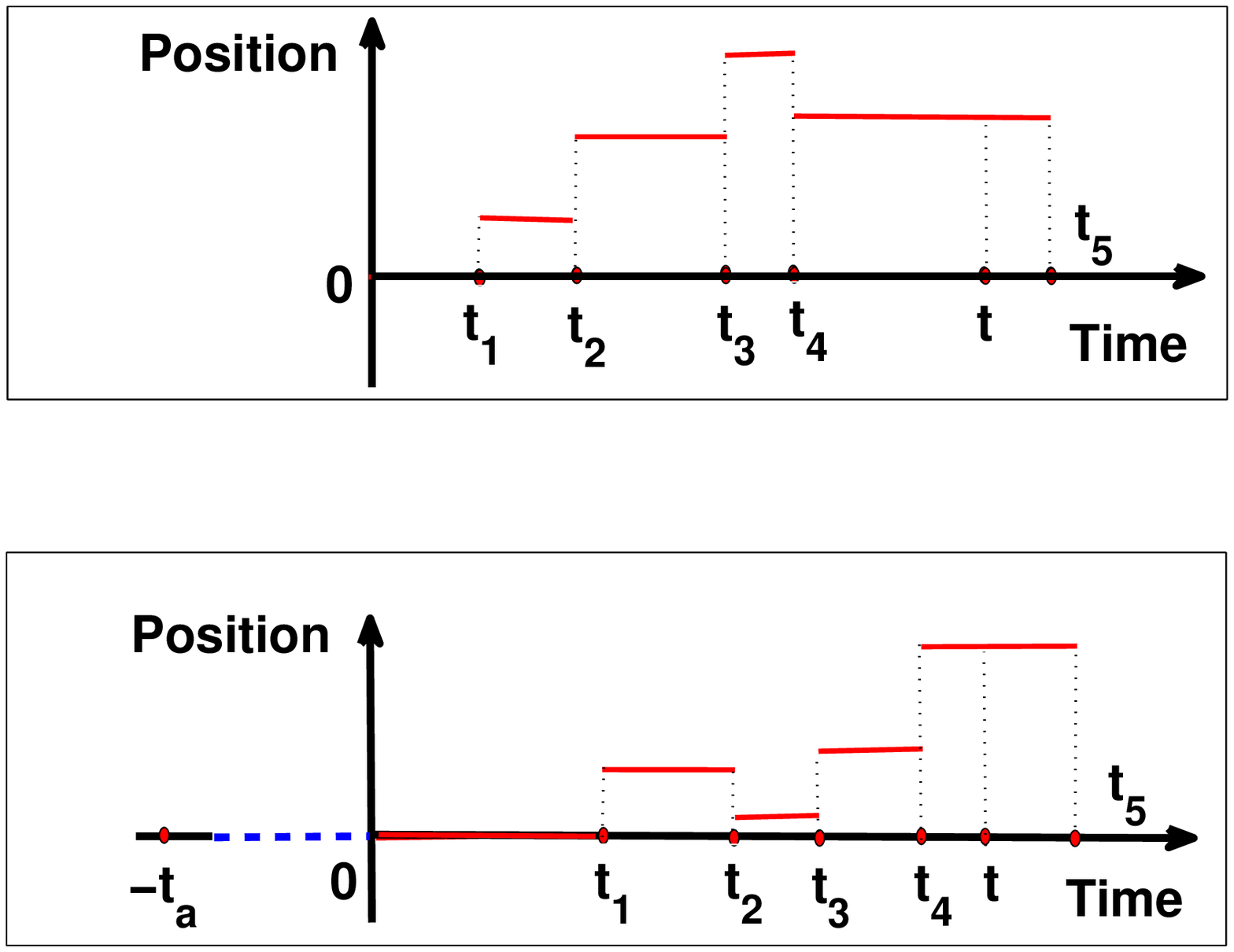}\\
  \caption{Illustrations of an ordinary CTRW (top) when the process begins at time $t=0$ and $x=0$, and an equilibrium CTRW (bottom).
The bottom panel describes an ongoing equilibrium process, i.e., a stationary case were the dynamics started long before the start of the observation (see the blue dashed line for an illustration).
The $t_i$ corresponds to the time when the $i$-th event occurs, and the backward recurrence time is $B_t=t-t_4$. The only difference between these two processes is the statistics of the waiting time of the first step. However due to the disorder, in particular the power-law trapping time distribution, this difference crucially influences the rare events and also the behavior of the MSD.
}
\label{trajectoryORvsEQ}
\end{figure}

The organization  of the paper is as follows. In Sec.~\ref{19sectsgb2}, we outline the single big principle and give the corresponding definitions.
Non-equilibrium and equilibrium initial conditions are investigated in Secs. \ref{18secttteess} and \ref{19sadeq100}, respectively.
Finally, we conclude the manuscript with a discussion. We also present  simulation results confirming the theoretical predications.

\section{Single big jump principle}\label{19sectsgb2}

\subsection{Model and definition}

We consider two types of biased CTRWs \cite{Metzler2000random,Barkai2003Aging,Klafter2011First,Schulz2014Aging,Kutner2017continuous}, the first is initiated at time
$t=0$ while the second is an equilibrium process. These two models, differ in the first trapping time statistics, but otherwise they are  identical. Let $\phi(\tau)$ be the probability
density function (PDF) of all the sojourn times while $h(\tau)$ is the PDF of the first one.
It should be  emphasized that the correct choice of $h(\tau)$ depends on the initial conditions.
For
the widely investigated non-equilibrium  initial condition, we assign  $h_{{\rm or}}(\tau)=\phi(\tau)$ \cite{Montroll1965Random}. This time process is sometimes called an ordinary renewal process, hence we use the subscript `or' to denote
this type of initial condition.  While
in equilibrium situation we use \cite{Haus1987Diffusion,Feller1971introduction,Tunaley1974Theory,Lax1977Renewal}
\begin{equation}\label{19wesgf120}
h_{{\rm eq}}(\tau)=\frac{\int_\tau^\infty\phi(y)dy}{\langle\tau\rangle},
\end{equation}
where $\langle\tau\rangle=\int_{0}^\infty \tau\phi(\tau)d\tau$ is the mean trapping time. We will soon explain the physical meaning
of these processes.

We are interested in the position of the random walker $x(t)$, which starts at $x=0$ when $t=0$. After waiting for time $\tau_1$, drawn from $h(\tau)$, the particle makes a spatial jump.
The PDF of jump size $\chi$, is Gaussian
\begin{equation}\label{18eq104}
  f(\chi)=\frac{1}{\sqrt{2\sigma^2\pi}}\exp\left[-\frac{(\chi-a)^2}{2\sigma^2}\right],
\end{equation}
where $a>0$ is the average size of the jumps.
Physically this is determined by an external constant force field that induces a net drift. From Eq.~\eqref{18eq104} the Fourier transform of $f(\chi)$ is $\widetilde{f}(k)=\exp(ika-\sigma^2 k^2/2)$. This yields
\begin{equation}\label{18eq105}
\widetilde{f}(k)\sim1+ika-\frac{\sigma^2+a^2}{2}k^2
\end{equation}
with $k\to 0$.
After the jump, say to $x_1$, the particle will pause for time $\tau_2$, whose statistical properties are drawn from $\phi(\tau)$. Then the process is renewed. We consider the widely applicable case,
where the PDF of trapping times  is
\begin{equation}\label{phit}
\phi(\tau)=\left\{
          \begin{split}
            &0, & \hbox{$\tau<\tau_0$;} \\
            &\alpha\frac{\tau_0^\alpha}{\tau^{1+\alpha}}, & \hbox{$\tau\geq\tau_0$}
          \end{split}
        \right.
\end{equation}
with
$1<\alpha<2$. As well-known such a fat-tailed distribution yields a wide
range of anomalous behaviors. See \cite{Bouchaud1990Anomalous,Metzler2000random} for review on CTRW and further discussion on physical systems below.
From the Abelian theorem, the Laplace $\tau\to s$ transform of $\phi(\tau)$  is
\begin{equation}\label{phits}
\widehat{\phi}(s)\sim 1-\langle\tau\rangle s+ b_\alpha s^\alpha
\end{equation}
with $b_\alpha=(\tau_0)^\alpha|\Gamma(1-\alpha)|$ and $s\to 0$. The leading term is the
normalization condition.
We focus on $1<\alpha<2$, where the mean $\langle \tau \rangle$ of the waiting time is finite, but not the variance. The
term $s^\alpha$ comes from the long tail of the waiting times
(and it is responsible for the deviations from the normal behavior).
Specific values of $\alpha$ for a range of physical systems and models are given in \cite{Bouchaud1990Anomalous,Metzler2014Anomalous}.

For an equilibrium initial condition the rate of performing a jump is stationary in
the sense that for any time $t$ the average number of jumps is
\begin{equation}\label{19wesgf120a}
\langle N(t)\rangle=\frac{t}{\langle\tau\rangle},
\end{equation}
so the effective rate $1/\langle\tau\rangle$ is a constant. In contrast, for the ordinary renewal
process we have in the long time limit $\langle N(t)\rangle\sim t/\langle\tau\rangle$, hence for short times the
two processes are not identical. Since the mean $\langle\tau\rangle$ is finite, one would expect
naively that statistical laws for the two processes will be identical in the long
time limit. While this is correct for some observables, for others this is false.
The prominent example is the MSD. In particular, for the
calculation of the rare events one must make the distinction between the two
models; see below.
Equilibrium initial condition is found when the particle is inserted in the medium
long before the process begins. More specifically when the process starts at
some time $-t_a$ before the measurement begins at time $t=0$, and in the limit
$t_a\to\infty$. All along we consider the displacement of the particle  compared to its initial position, namely we assign $x(0)=0$. For a schematic presentation of the random processes see Fig.~\ref{trajectoryORvsEQ}.

Non-equilibrium initial conditions are found when the processes are initiated at time $t=0$. For example in Scher Montroll theory \cite{Scher1975Anomalous}, a flash of light excites
charge carriers at time $t=0$ and then the process of diffusion begins, then we
have an ordinary process.
Mathematically, these two models  merely differ by the statistics of  the waiting time of the first step, and hence it is interesting to compare them,
to see if this seemingly small modification of the model is important or not in
the long time limit. For a Poisson process the two models are identical. In contrast, for heavy-tailed processes under investigation, we find from Eqs. \eqref{19wesgf120} and \eqref{phit} that
\begin{equation}\label{friswairss}
h_{{\rm eq}}(\tau)\sim \frac{(\tau_0)^\alpha}{\langle\tau\rangle}\tau^{-\alpha}.
\end{equation}
As $1<\alpha<2$ we see that the average time  for the first waiting time
diverges (but not for the second etc).
This means that in a stationary state the process is slower if compared to the ordinary case, hence we expect that in this case particles will be lagging even more behind the mean displacement.

Let us discuss the applicability of the CTRW model. As mentioned Scher and Montroll showed how this theory describes diffusion of charge carriers in disordered medium. In some experiments, one can find $\alpha=T/T_g$ where $T$ is the temperature and $T_g$ is the measure of the disorder. This is also the case for Bouchaud trap model describing glassy dynamics \cite{Bouchaud1990Anomalous}. In the context of contamination spreading biased CTRW is used with $\alpha=1.73$ \cite{Alon2017Time}. Based on numerical simulations, Winter and Schroer showed  the super diffusive behavior  and related the dynamics to the biased CTRW \cite{Winter2012Active,Schroer2013Anomalous}. In these systems one expects that at very long times we find normal diffusion.  There are also many examples of CTRW without bias \cite{Metzler2000random,Metzler2014Anomalous,Krapf2016Strange,Edery2018Surfactant}. It is interesting to add a bias in these systems to compare the effect discussed here.


\subsection{Main results: the big jump principle}
 \begin{figure}[htb]
  \centering
  \includegraphics[width=9cm, height=6cm]{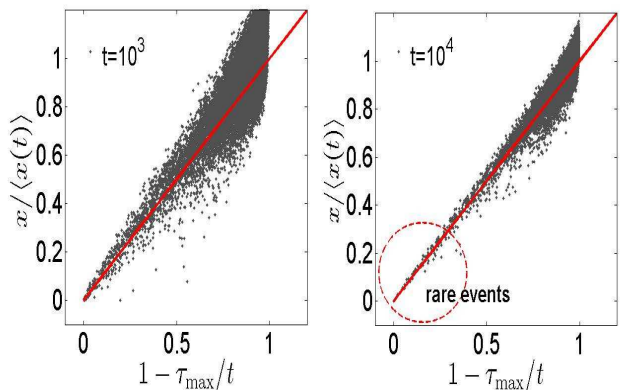}\\
  \caption{ A correlation plot between $1-\tau_{\max}/t$ and $x/\langle x(t)\rangle$ predicated in Eq.~\eqref{19seq2} for the biased ordinary  CTRW process. Here  we choose  $a=1$, $\alpha=1.5$, $\sigma=\sqrt{2}$, and $\langle\tau\rangle=1$. The dots  are simulation results obtained by generating $10^5$ trajectories and the red solid line is obtained from Eq.~\eqref{19seq2} by switching random variables to a dimensionless form, i.e., $x/\langle x(t)\rangle\sim x/(at/
  \langle\tau\rangle) \doteqdot (1-\tau_{\max}/t)$.
The evidently strong correlations, circled on the right panel, indicate that a single trapping event is responsible for the statistics of rare events.
 }
\label{PositionVsTauRare}
\end{figure}
Transport and diffusion processes, either normal or anomalous, are composed of a large number of displacements. Hence statistical laws, like the central limit theorem, are useful tools describing universal aspects of the phenomenon. In our case a single event is controlling the statistics of the spreading packet at its tail. Let $\{\tau_1,\tau_2,\cdots,\tau_N,B_t\}$ be the set of the waiting times between jump events, and $\sum_{i=1}^{N}\tau_i+B_t=t$ is the measurement time. Here $B_t$, called the backward recurrence time, is the time elapsing between the moment of last jump $t_N=\sum_{i=1}^N\tau_i$ and the measurement time $t$. $N$ is the random number of jumps in $(0,t)$ \cite{Godreche2001Statistics}. We define the largest waiting time according to
 \begin{equation}\label{19seq1}
 \tau_{\max}=\max\{\tau_1,\tau_2,\cdots,\tau_N,B_t\}.
 \end{equation}
One main conclusion of this manuscript is that the statistics of $\tau_{\max}$ determines the fluctuations of the position $x(t)$ of the biased random walker. This holds for rare fluctuations of $x(t)$, that still control the behavior of the most typical observable in the field: the MSD.

Due to the fat tailed distribution of the trapping time $\phi(\tau)$, and using basic arguments from extreme value statistics of  IID random variables, one expects that the typical fluctuations scale like $\tau_{\max}\propto t^{1/\alpha}$, while for a thin tailed distribution of waiting time, e.g., $\phi(\tau)=\exp(-\tau)$, we have $\tau_{\max}\propto \log(t)$ \cite{Gumbel2004Statistics}. For the latter example `$\propto$' means that $\tau_{\max}$ is of the order of $\log(t)$ and similarly for the former case. While we are not dealing with IID random variables, the constraint is weak in the sense that it does not modify the typical fluctuations, see below and Ref. \cite{Cistjakov1964theorem,Godreche2015Statistics}.
Note that all these scalings, i.e., $\tau_{\max}\propto t^{1/\alpha}$ and $\tau_{\max}\propto \log(t)$, describe typical fluctuations, sometimes called bulk fluctuations. These fluctuations are described by normalized densities, specified by Fr\'{e}chet's law and the Gumbel law.
On the contrary, here we focus on rare fluctuations, that is to say, both  $\tau_{\max}$ and  $t$ are comparable.

When Eq.~\eqref{phit} holds, for the biased CTRW we will demonstrate that for small $x$, i.e., the left plume in Fig.~\ref{trajectory}
 \begin{equation}\label{19seq2}
x \doteqdot \frac{t-\tau_{\max}}{\langle\tau\rangle} a,
 \end{equation}
where ``$\doteqdot$'' indicates that the random variables on both sides follow the same distribution. However, the PDFs describing  the position of the particle $x$ when $x$ is not small
and of $\tau_{\max}$ are far from being identical, indeed they will be calculated below.
The meaning of small $x$ and large $\tau_{\max}$ will soon become clear when we
formulate the problem more precisely. For now based on Figs.~\ref{trajectory} and \ref{trajectoryv2}, we see Eq.~\eqref{19seq2}  works well when $x\ll \langle x(t)\rangle$ and $\tau_{\max} \simeq  t$. For example
when $x<50\ll \langle x(t)\rangle\simeq1667$ in the bottom panel of Fig. \ref{trajectory}, or for the trajectory on the left panel of Fig.~\ref{trajectoryv2}, where $\tau_{\max}=988$, when $t=1000$ and then $x\simeq40\ll\langle x(t)\rangle\simeq1667$.
%
%
%
Eq.~\eqref{19seq2} means that the distribution of  $x\ll\langle x(t)\rangle$ is the same as the average size of the jumps  $a$, times the typical number of jumps made in $(0,t-\tau_{\max})$, which is the time `free' of the longest waiting time. A correlation plot based on Eq.~\eqref{19seq2} is demonstrated numerically in Fig.~\ref{PositionVsTauRare}. Using simulations of the ordinary CTRW process,  we generate trajectories and search for positions of the random walkers at time $t$ and record $\tau_{\max}$. Then we plot the random  entries observing that  for small $x$, there is a perfect correlation as predicated by Eq.~\eqref{19seq2}.
Such correlation plots indicate that  Eq.~\eqref{19seq2} is working.
We call this the principle of big jump, and it is valid  for both stationary and  ordinary processes. Here the big jump means large trapping time, see further discussion on the term big jump and its origin in the discussion and  summary.
Now we will   analytically derive
 Eq.~\eqref{19seq2}  and discuss its consequence.
 For that we obtain the distribution of  $\tau_{\max}$ and then of $x$.

\begin{remark}
Our main results in this manuscript are Eqs.~\eqref{19seq4}, \eqref{19seq13}, \eqref{19sadeq102fg}, and \eqref{19alwSec3qw1010} which give explicit formulas for the PDF of $x$ and $\tau_{\max}$ for the two types of processes under  investigation.
In \cite{Alessandro2019Single} we promoted a rate formalism to treat similar problems, e.g. the L\'{e}vy walk.
Here the focus is on
the exact calculation of the statistics of
rare events both for  $\tau_{\max}$ and
$x$, and on the relation between
these two random variables, i.e.,
Eq. \eqref{19seq2}.

\end{remark}

\subsection{The statistics of $\tau_{\max}$}
Let us proceed to derive the general  formulas describing the statistics of the longest waiting times which are valid for both  the ordinary and equilibrium renewal processes. The case of an ordinary renewal theory, was considered
previously  by Godr{\'e}che, Majumdar and  Schehr in Ref. \cite{Godreche2015Statistics}. They investigated the typical fluctuations of $\tau_{\max}$, and these as explained below exhibit behavior identical to a classical case of extreme value statistics, namely Fr\'{e}chet's law holds for typical fluctuations. Here our goal is very different, we aim to obtain the rare events, namely investigate the behavior when $\tau_{\max}$ is of the order $t$. In this case the fluctuations greatly differ from the IID case.

We define the probability that $\tau_{\max}$ is smaller than $L$
\begin{equation}\label{weesdfesstyj}
\begin{split}
 F(t,L) &=Prob[\tau_{\max}\leq L].
\end{split}
\end{equation}
The corresponding PDF is $P_{\tau_{\max}}(t,L)$ and as usual $F(t,L)=\int_0^LP_{\tau_{\max}}(t,y)dy$.
Clearly the probability depends on the measurement time $t$ and this dependence is especially important for fat-tailed waiting time PDFs.
It is helpful to introduce the joint probability  distribution of $\tau_{\max}$ and the number of renewals $N$
\begin{widetext}
\begin{equation}\label{19alwSec3qw101}
\begin{split}
  F_{n}(t,L) & =Prob(\tau_{\max}\leq L,N=n) \\
    &=\int_0^Ld\tau_1 \int_0^Ld\tau_2\cdots\int_0^Ld B_t P_{\tau_{\max}}(\tau_1,\tau_2,\cdots,\tau_n,B_t)\\
    &=\int_0^Lh(\tau_1) d\tau_1\int_0^L\phi(\tau_2)d\tau_2\cdots\int_0^L\Phi(B_t)dB_t\delta\left(t-\left(\sum_{1}^{n}\tau_j+B_t\right)\right).
\end{split}
\end{equation}
Here $h(\cdot)$ in the third line of Eq.~\eqref{19alwSec3qw101} is governed by the process we investigate, and $\Phi(t)$ determined by  the type of the process and the number of renewals
\begin{equation}\label{19smeq3}
\displaystyle \Phi(t)=\left\{
         \begin{array}{ll}
          \displaystyle \int_t^\infty h(\tau)d\tau, & \hbox{$n=0$, equilibrium process;} \\
          \displaystyle \int_t^\infty\phi(\tau)d\tau, & \hbox{otherwise.}
         \end{array}
       \right.
\end{equation}
Taking the Laplace transform w.r.t. $t$, we find

\begin{equation}\label{19alwSec3qw102}
  \widehat{F}_{n}(s,L)=\left\{
                   \begin{array}{ll}
                    \displaystyle \int_0^L\exp(-s\tau_1)\int_{\tau_1}^\infty h(\tau)d\tau d\tau_1, & \hbox{$n=0$;} \\
                \displaystyle  \int_0^L\exp(-s\tau)h(\tau_1) d\tau_1\left(\int_0^L\exp(-s\tau)\phi(\tau)d\tau\right)^{n-1}\int_0^L\exp(-sB_t)\int_{B_t}^\infty \phi(\tau)d\tau dB_t, & \hbox{$n\geq 1$.}
                   \end{array}
                 \right.
\end{equation}
The case $n=0$ corresponds to realizations with no renewals during the time interval $(0,t)$.
One can check that $\sum_{n=0}^\infty\widehat{F}_{n}(s,L\to \infty)=1/s$.
This means that the density of $\tau_{\max}$ is normalized.
The sum of $n$ from zero to infinity gives
\begin{equation}\label{19alwSec3qw103}
  \widehat{F}(s,L)=\int_0^L\exp(-s\tau_1)\int_{\tau_1}^\infty h(\tau)d\tau d\tau_1+ \int_0^L\exp(-s\tau_1)h(\tau_1) d\tau_1\frac{\int_0^L\exp(-sB_t)\int_{B_t}^\infty\phi(y)dyd{B_t}}{1-\int_0^L\exp(-s\tau)\phi(\tau)d\tau}.
\end{equation}
The first term is related to the survival probability and the second term corresponds to the probability that at least one renewal happened in $(0,t)$. For the equilibrium renewal process, we insert Eq.~\eqref{19wesgf120} into Eq.~\eqref{19alwSec3qw103} while for the ordinary case we use $h_{{\rm or}}(\tau)=\phi(\tau)$.
Below, from Eq.~\eqref{19alwSec3qw103} we will calculate the far tail of the distribution of $\tau_{\max}$ for the two different processes, i.e., the ordinary process and the equilibrium one, and prove that Eq.~\eqref{19seq2} is valid for both cases.
\end{widetext}

\section{The ordinary process}\label{18secttteess}
Here we   consider the ordinary renewal process and the ordinary CTRW to build the relation between the rare events of positions and the largest waiting times.
\begin{figure}[htb]
 \centering
 \includegraphics[width=8cm, height=5.5cm]{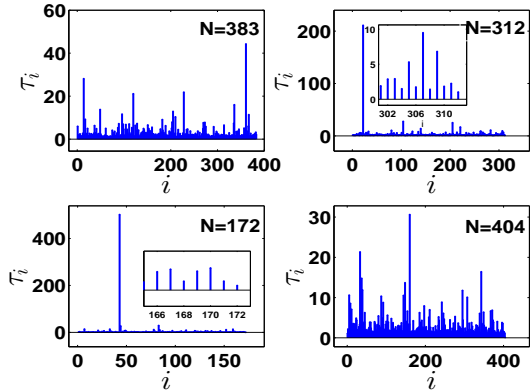}\\
 \caption{Random variables $\tau_i$ for an ordinary renewal process with Eq.~\eqref{phit} and $\alpha=1.5$. The observation time $t$ is $1000$ and $i=100$, $200$, $\cdots$ correspond to the $100$-th, $200$-th, $\cdots$ waiting time of the fractional renewal process, respectively. Clearly  in our case the number of  renewals, shown in the top-right corner of the subplots, is a random variable; see inset. Due to the
fat-tailed trapping times, the fluctuations of $N$ is large which comes from the large fluctuations of waiting times.
}\label{WAITINGTIMERandomN}
\end{figure}

\subsection{The rare fluctuations of $\tau_{\max}$}\label{howtaumaxnoor}
The aim is to investigate the PDF of $\tau_{\max}$ for the non-equilibrium process which is  denoted as $P_{{\rm or},\tau_{\max}}(t,L)$.
We first treat the problem heuristically  to calculate the typical fluctuations. Let $\langle N\rangle=t/\langle\tau\rangle$ be the average number of renewals in the long time limit. For simplification, we neglect $B_t$  in Eq.~\eqref{19seq1} and ignore the constraint $\sum_{i=1}^N\tau_i+B_t=t$, further we  replace the random $N$ with $\langle N\rangle$. This means that we treat this problem as if the waiting times are IID random variables, an approximation which turns out not sufficient in our case, still ignoring the correlation \cite{Godreche2015Statistics}
 \begin{equation}\label{19seq3}
 \begin{split}
  Prob(\tau_{\max}<L)&=Prob^{N}(\tau_i<L)\\
  &\simeq \left[1-\left(\frac{\tau_0}{L}\right)^\alpha\right]^N\\
  &\sim \exp\left[-\langle N\rangle \left(\frac{\tau_0}{L}\right)^\alpha\right].
 \end{split}
 \end{equation}
 This is the well-known Fr{\'e}chet distribution \cite{Albeverio2006Extreme}.
A closer look reveals  a drawback of this treatment of the typical fluctuations, since within this approximation the PDF of $\tau_{\max}$ is  $P_{\tau_{\max}}(t,L)\sim \alpha\langle N\rangle(\tau_0)^\alpha/L^{1+\alpha}$, for $L\to \infty$.
However in our  setting $\tau_{\max}\leq t$. This means that we must modify  Fr{\'e}chet's law at its tail, in other words, the constraint that the sum of all the waiting times and the backward recurrence time is equal to the measurement time $t$,   comes into play when $\tau_{\max}\propto t$,  as expected. Note that the number of renewals in our case is a random variable; see Fig.~\ref{WAITINGTIMERandomN}.

\begin{figure}[h]
 \centering
 \includegraphics[width=9cm]{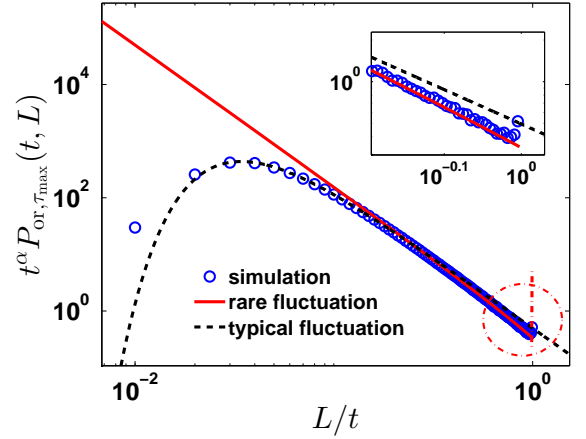}\\
 \caption{ Scaled PDF of the longest time interval $t^{\alpha}P_{{\rm or},{\tau_{\max}}}(t,L)$ versus $L/t$. The red solid lines predicated by Eq.~\eqref{19smeq9} or equivalently  Eq.~(\ref{19seq4}), describe the rare fluctuations showing the behavior of $L$ when it is of the order of $t$.
The simulations, plotted by the symbols, are generated by averaging $10^6$ trajectories with $\alpha=3/2$ and $\tau_0=1$. The figure clearly shows a perfect agreement of the simulations  compared with the theoretical result Eq.~\eqref{19smeq9}, which  has a sharp cutoff at the tail of density at $\tau_{\max}/t\to 1$ (see the red dash-doted lines). This is very different if compared with   typical fluctuations calculated with  Fr{\'e}chet's distribution Eq.~(\ref{19seq3}), which clearly does not describe well
the far tail (see inset).
}\label{TauMaxRare}
\end{figure}

Now we use an exact solution of the problem to calculate the rare events.
Considering the non-equilibrium renewal  process, we insert $h(\tau)=\phi(\tau)$ into Eq. \eqref{19alwSec3qw102}  to get \cite{Godreche2015Statistics}
\begin{equation}\label{2019lgsec2a01}
\int_{L}^{\infty} \widehat{P}_{{\rm or},{\tau_{\max}}}(z)dz=\frac{1}{s}\frac{1}{1+\widehat{G}(s,L)}
\end{equation}
with
\begin{equation}\label{19smeq1ggggg}
\widehat{G}(s,L)=\frac{s\exp(sL)}{p_0(L)}\int_0^{L}p_0(t)\exp(-st)dt
\end{equation}
and  the survival probability
\begin{equation}\label{19smeq1}
p_0(t)=\int_t^\infty\phi(\tau)d\tau\simeq \left(\frac{\tau_0}{t}\right)^\alpha.
\end{equation}
We are interested in the limit $s\to 0$ (corresponding to
long measurement time) and $L\to \infty$ in such a way
that $sL$ remains a constant. As mentioned the typical fluctuations  are described by Fr\'{e}chet's law Eq.~\eqref{19seq3}
and here instead we consider the rare fluctuations. Using Eq.~\eqref{19smeq1}, for $L\to\infty$, Eq.~\eqref{19smeq1ggggg} becomes
\begin{equation}\label{19smeq7}
\begin{split}
  \widehat{G}(s,L) 
    &\sim \frac{\exp(sL)sL^\alpha\langle\tau\rangle}{(\tau_0)^\alpha},
\end{split}
\end{equation}
where we have used the limit
\begin{equation}\label{19smeq701}
\lim_{L\to \infty}\int_0^{L} p_0(t)\exp(-st)dt=\frac{1-\widehat{\phi}(s)}{s}
\sim \langle\tau\rangle
\end{equation}
 with $s\to 0$.
From Eq.~\eqref{19smeq7} we see  that $\widehat{G}(s,L)$ is large for $sL\to \rm{constant}$ and $\alpha>1$.
According to Eq.~\eqref{19smeq7}, we find
\begin{equation}\label{19sadd01}
\frac{\partial \widehat{G}(s,L) }{\partial L}\sim s\widehat{G}(s,L)+\frac{\alpha}{L} \widehat{G}(s,L)+\cdots.
\end{equation}
Note that Eq.~\eqref{19sadd01} can also be derived directly from Eq.~\eqref{19smeq1ggggg}.
Utilizing Eq.~\eqref{2019lgsec2a01} and
\begin{equation}\label{19smeq2}
F(t,L)=\int_0^{L}P_{{\rm or},\tau_{\max}}(t,y)dy,
\end{equation}
and after some simple rearrangements
\begin{equation}\label{19smeq5}
\widehat{P}_{{\rm or},{\tau_{\max}}}(s,L)=\frac{1}{s}\frac{\frac{\partial \widehat{G}(s,L)}{\partial L}}{[1+\widehat{G}(s,L)]^2},
\end{equation}
where we used the relation that $P_{{\rm or},\tau_{\max}}(t,L)$ is the derivative of Eq.~\eqref{19smeq2} w.r.t. $L$.
Combining  Eqs.~\eqref{19sadd01} and \eqref{19smeq5}, we have
\begin{equation}\label{19smeq6}
\widehat{P}_{{\rm or},{\tau_{\max}}}(s,L)\sim\frac{1}{\widehat{G}(s,L)}+\frac{\alpha}{sL\widehat{G}(s,L)}+\cdots.
\end{equation}
Note that the first two terms on the right-hand side of Eq.~\eqref{19smeq6}, namely $1/\widehat{G}(s,L)$ and $\alpha/(\widehat{G}(s,L)sL)$, are comparable when $sL\to \rm{constant}$. Hence from Eqs.~\eqref{19smeq7} and \eqref{19smeq6}, we get
\begin{equation}\label{19smeq8}
\widehat{P}_{{\rm or},{\tau_{\max}}}(s,L)\sim \frac{(\tau_0)^\alpha}{\langle\tau\rangle}\frac{\exp(-sL)}{sL^\alpha}\left(1+\frac{\alpha}{sL}     \right).
\end{equation}
Taking the inverse Laplace transform $s\to t$ of Eq.~\eqref{19smeq8} gives our second main result with
 the scaling $L\propto t$
\begin{equation}\label{19smeq9}
  P_{{\rm or},{\tau_{\max}}}(t,L)\sim   \frac{(\tau_0)^\alpha}{t^\alpha\langle\tau\rangle} \left[\alpha\left(\frac{L}{t}\right)^{-\alpha-1}
    -(\alpha-1)\left(\frac{L}{t}\right)^{-\alpha}\right]
\end{equation}
with $0\leq L\leq t$.  Theoretical predication of Eq.~\eqref{19smeq9} is compared with numerical simulations in Fig.~\ref{TauMaxRare}.
As explained before, Eq.~\eqref{19smeq9} describing the far tail of the distribution of $\tau_{\max}$ is a modification of Fr{\'e}chet's law.

According to Eq.~\eqref{19smeq9}, we find

\begin{equation}\label{19seq4}
\lim_{t\to\infty}\langle\tau\rangle\left(\frac{t}{\tau_0}\right)^\alpha  P_{{\rm or},{\tau_{\max}}}(t,L)= \mathcal{I}_{{\rm or},\alpha}\left(\frac{L}{t}\right),
\end{equation}
where
\begin{equation}\label{19seq4add}
\mathcal{I}_{{\rm or},\alpha}(y)=\alpha y^{-\alpha-1}-(\alpha-1)y^{-\alpha}
\end{equation}
with $0<y<1$.
This scaling solution describes the far tail of the distribution, where Fr{\'e}chet's law does not work. In fact, these two laws are related as the $y^{-\alpha-1}$ term matches the far tail of  the Fr{\'e}chet law, as it should. Since $0<y<1$ implies $\tau_{\max}<t$,  moments of $\tau_{\max}$ are computed w.r.t. this scaling solution. In contrast, the Fr{\'e}chet law  gives diverging variance of $\tau_{\max}$, which is certainly not a possibility since $\tau_{\max}$ is bounded.
The expression in Eq.~\eqref{19seq4} is an infinite
density  describing  a non-normalising limiting law. More exactly $\mathcal{I}_{{\rm or},\alpha}(\cdot)$ is not normalizable, the moments of order $q>\alpha$  of $\tau_{\max}$ are calculated w.r.t. this non-normalised state. More details on infinite densities see Refs.~\cite{Aaronson1997introduction,Kessler2010Infinite,Rebenshtok2014Infinite,Wang2018Renewal,Erez2019From}

\subsection{The rare fluctuations of the position}

We now investigate the distribution of $x$ proving the validity of the big jump principle  Eq.~\eqref{19seq2}. Let $P_{{\rm or}}(x,t)$ be the PDF of finding the walker on $x$ at time $t$. The starting point is the well-known Montroll-Weiss equation, which gives Fourier-Laplace transform of
the $P_{{\rm or}}(x,t)$
 \cite{Bouchaud1990Anomalous,Metzler2000random}
\begin{equation}\label{19seq5}
\widetilde{\widehat{P}}_{{\rm or}}(k,s)=\frac{1-\widehat{\phi}(s)}{s}\frac{1}{1-\widetilde{f}(k)\widehat{\phi}(s)}
\end{equation}
with $\widetilde{\widehat{P}}_{{\rm or}}(k,s)=\int_{-\infty}^{\infty}\int_0^{\infty}\exp(ikx-st)P_{{\rm or}}(x,t)dtdx$.
Here $\widetilde{f}(k)$ is the Fourier transform of the jump length PDF $f(\chi)$, and $\widehat{\phi}(s)$ is the Laplace transform of waiting time PDF. The long wave length approximation, i.e., the small $s$ and $k$ limit, is routinely applied to investigate the long time limit  of $P_{{\rm or}}(x,t)$. However, how to choose the limit of $k\to 0$ and $s\to 0$ is actually slightly subtle. Utilizing Eqs.~\eqref{18eq105} and \eqref{phits}, and  assuming  that the ratio $|s^\alpha|/|k|$ is fixed, we get
\begin{equation}\label{19seq6}
\widetilde{\widehat{P}}_{{\rm or}}(k,s)\sim\frac{\langle\tau\rangle}{-ika+s\langle\tau\rangle-(\tau_0)^\alpha|\Gamma(1-\alpha)| s^\alpha},
\end{equation}
Inverting, we then find a known limit theorem \cite{Kotulski1995Asymptotic,Burioni2013Rare}
\begin{equation}\label{19seq7}
P_{{\rm or}}(x,t)\sim\frac{1}{a(t/\overline{t})^{1/\alpha}}L_{\alpha,1}\left(\frac{x-at/\langle\tau\rangle}{a(t/\overline{t})^{1/\alpha}}\right),
\end{equation}
where $\overline{t}=\langle\tau\rangle^{1+\alpha}/((\tau_0)^\alpha|\Gamma(1-\alpha)|)$, $L_{\alpha,1}(\cdot)$ is the non-symmetrical L{\'e}vy stable law with characteristic function $\exp[(ik)^\alpha]$, and $a>0$. This central limit theorem, just like Fr{\'e}chet's law, has its limitations.  As a stand alone formula, it predicates $\langle x^2(t)\rangle=\infty$, since the second moment of the L{\'e}vy distribution does not exist. This means that we must consider a different method to describe the far tail.

To proceed we  reanalyze Eq.~\eqref{19seq5} but now fixing $|s|/|k|$. This is  a large deviation approach since such a scaling implies a  ballistic scaling behavior of $x$ and $t$, unlike $x-at/\langle\tau\rangle \propto t^{1/\alpha}$ in Eq.~\eqref{19seq7}.
The strategy we use now, i.e., the determination of $P_{{\rm or}}(x,t)$ for $x\propto t$, is similar to the approach in the previous section  where we calculated  $P_{{\rm or},\tau_{\max}}(t,L)$. The obvious difference is that there we start  with Eq.~\eqref{2019lgsec2a01}, while here with the Montroll-Weiss Eq.~\eqref{19seq5}. More specifically in the Sec. \ref{howtaumaxnoor} we assume that $sL\propto \rm{constant}$, while here $|s|$ and $|k|$ are small and of the same order, where $s$ and $k$ are Laplace pair and Fourier pair of $t$ and $x$, respectively.

\begin{figure}[h]
  \centering
  \includegraphics[width=9cm]{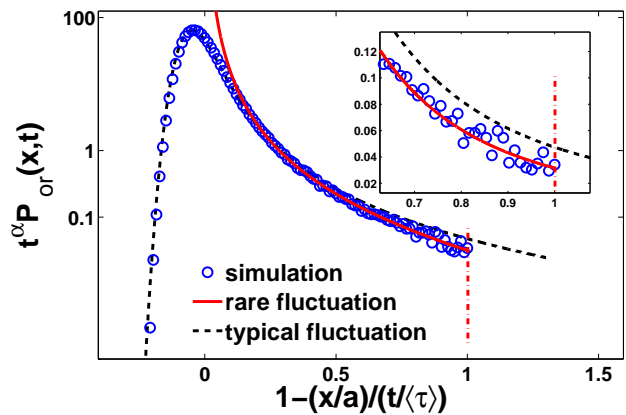}\\
  \caption{
  Scaled PDF $P_{{\rm or}}(x,t)$ is compared with the prediction of the single big jump principle  and the L{\'e}vy central limit theorem describing rare and typical fluctuations.
The parameters are $a=1$, $\sigma=1$, $\alpha=1.5$, $\tau_0=0.1$, and for the simulation we used $5\times10^6$ trajectories.
The inset exhibits a comparison among typical fluctuations Eq. \eqref{19seq7}, rare fluctuations Eq. \eqref{19seq13}, and simulations.
Clearly our theory performs perfectly, while Eq.~\eqref{19seq7} over shoots (see inset) and extents
to the positive infinity. In reality there is  a clear cutoff at $x=0$ being exclusively revealed by the single big jump principle
analysis (see the marked red dash-doted lines).
}
\label{PxVsxiVsTimeRare15}
\end{figure}

We restart from Eq.~\eqref{19seq5}, which gives
\begin{equation}\label{19SEQ8}
\widetilde{\widehat{P}}_{{\rm or}}(k,s)\sim
\underbrace{\frac{\langle\tau\rangle}{\langle\tau\rangle  s-ika}}_{{ {\rm leading}}}+
\underbrace{\frac{ika (\tau_0)^\alpha|\Gamma(1-\alpha)| s^{\alpha-1}}{(s\langle\tau\rangle-ika)^2}}_{{ {\rm correction}}}+\cdots.
\end{equation}
The derivation of Eq.~\eqref{19SEQ8} is given in Appendix \ref{19smss2}.
The inversion of the leading term is trivial, but it yields a delta function  $\delta(x-at/\langle\tau\rangle)$. Mathematically we choose a scaling that shrinks the density to an uninteresting object. Luckily, the correction term is important as it describes the far tail. So for $x\neq at/\langle\tau\rangle$, we have
\begin{equation}\label{19seq9}
P_{{\rm or}}(x,t)\sim \mathcal{F}^{-1}_{k\to x}\mathcal{L}^{-1}_{s\to t}\left[\frac{a(\tau_0)^\alpha|\Gamma (1-\alpha)| iks^{\alpha-1}}{(s\langle\tau\rangle-ika)^2}\right]
\end{equation}
with $\mathcal{F}^{-1}_{k\to x}$ and $\mathcal{L}^{-1}_{s\to t}$ being the inverse Fourier and the inverse Laplace transforms, respectively.
We first perform the inverse Laplace transform using the convolution theorem and the pairs
\begin{equation}\label{19seq10}
\left\{
  \begin{array}{ll}
    \displaystyle  \mathcal{L}^{-1}_{s\to t}[s^{\alpha-1}]=\frac{t^{-\alpha}}{\Gamma(1-\alpha)}, & \hbox{~} \\
   \displaystyle  \mathcal{L}^{-1}_{s\to t}\left[\frac{1}{(s-ika/\langle\tau\rangle)^2}\right]=t\exp\left(ika\frac{t}{\langle\tau\rangle}\right) & \hbox{~}
  \end{array}
\right.
\end{equation}
and find
\begin{equation}\label{19seq11}
P_{{\rm or}}(x,t)\sim \mathcal{F}^{-1}_{k\to x}\left[-ik\frac{a(\tau_0)^\alpha}{\langle\tau\rangle^2}\int_0^t \frac{y\exp\left(\frac{ikay}{\langle\tau\rangle}\right)}{(t-y)^{\alpha}} dy
\right].
\end{equation}
The inverse Fourier transform of $\exp(ikay/\langle\tau\rangle)$ is a delta function and the $ik$ in front of this expression is the spatial derivative in $x$ space, hence we get
\begin{equation}\label{19seq12}
P_{{\rm or}}(x,t)\sim \frac{(\tau_0)^\alpha  }{\langle\tau\rangle}\frac{\partial}{\partial x}\int_0^t\frac{y\delta\left(y-\frac{x\langle\tau\rangle}{a}\right)}{(t-y)^{\alpha}}
dy.
\end{equation}
Then after simple rearrangements
\begin{equation}\label{19seq13}
P_{{\rm or}}(x,t)\sim \frac{(\tau_0)^\alpha }{at^{\alpha}}\mathcal{I}_{{\rm or},\alpha}(\xi)
\end{equation}
with $0<\xi<1$, $\xi=1-(x/a)/(t/\langle\tau\rangle)$, and $\mathcal{I}_{{\rm or},\alpha}(\cdot)$ being defined by Eq.~\eqref{19seq4add}.   As  Fig.~\ref{PxVsxiVsTimeRare15} demonstrates,    this equation describes the far tail of the density of the spreading packet, and it is complementary to the L{\'e}vy law  Eq.~\eqref{19seq7}. The MSD of the process is obtained w.r.t. integration over the formula Eq.~\eqref{19seq13} and in that sense this equation ``cures'' the drawback of the L{\'e}vy density. More importantly is the fact that the distribution of $\tau_{\max}$ Eq.~\eqref{19seq4} and $x$ Eq.~\eqref{19seq13} have the same structure, beyond a trivial Jacobian. In other words, given the fact that these observables have the same distribution, we have proven the single big jump principle Eq.~\eqref{19seq2} for the ordinary processes.  The statistics of one waiting time, $\tau_{\max}$, determines the fluctuations at small $x$. And since Eq.~\eqref{19seq13} gives the MSD, which  is used in most experimental, theoretical and numerical works to characterize the fluctuations, we see that the MSD is directly related to the single big jump principle and extreme value statistics. One should note that low-order moments like $\langle |x-\langle x\rangle|^q\rangle$ with $q<\alpha$ are finite w.r.t. the L{\'e}vy density, and these are given by integration w.r.t. Eq.~\eqref{19seq7}.

\begin{remark}
We now study the case of CTRW in two dimensions and focus on an ordinary process. The joint length PDF is $f(\chi_x,\chi_y)=f_x(\chi_x)f_y(\chi_y)$ where $f_x(\chi_x)$ is the same as in Eq.~\eqref{18eq104} and $f_y(\chi_y)=\exp(-(\chi_y)^2/2(\sigma_y)^2)/(\sqrt{2\pi(\sigma_y)^2})$ with $\sigma_y$ being a constant. This means that the drift is only  in $x$ direction. Similar to our previous calculations, we use the Montroll-Weiss equation and find
\begin{equation}\label{foote3}
P_{\rm{or}}(x,y,t)\sim \frac{(\tau_0)^\alpha }{at^{\alpha}}\mathcal{I}_{{\rm or},\alpha}(\xi)\delta(y).
\end{equation}
The marginal density $P_{{\rm or}}(x,t)$ is the same as  the one dimensional case Eq.~\eqref{19seq13}. Note that $\tau_{\max}$ is of the order $t$ (for the far tail), so in the $y$ direction the particles are practically frozen. Hence we get a delta function since there is no drift in the $y$ direction.
\end{remark}

\section{The equilibrium case}\label{19sadeq100}
Up to now we have considered the case when a physical clock  was started immediately at the beginning of the process, i.e., an ordinary  CTRW. Here we  consider the equilibrium initial condition. We note that for $0<\alpha<1$, i.e., when the average trapping time diverges, this is related to Aging CTRW \cite{Schutz1997Single,Kues2001Visualization,Barkai2003Aging,Metzler2014,Schulz2014Aging} which is used as a tool to describe complex systems ranging from  Anderson insulator to colloidal
suspensions and it was first introduced by Monthus and Bouchaud to illustrate the diffusion in glasses \cite{Monthus1996Models}.  In contrast, when $1<\alpha<2$ and
Eq. \eqref{19wesgf120} holds, we have a stationary process.
Then as mentioned already, the mean waiting time for the first event is infinite; see Eq.~\eqref{friswairss}.
In practice, if we start the process at time $-t_a$ and $t_a$ is large but finite the
averaged first waiting time observed
after time $t_a$ will increase with $t_a$, and when $t_a$ tends to infinity it will diverge.
Here we focus on the statistics of particles  with an equilibrium condition, i.e., $t_a\to\infty$.

\subsection{The rare fluctuations of the position}
In Fourier-Laplace space, the density of spreading particles is given by \cite{Barkai2003Aging}
\begin{equation}\label{19sadeq101F1addf}
\begin{split}
  \widetilde{\widehat{P}}_{\rm{eq}}(k,s) & =\frac{1-\widehat{h}_{{\rm eq}}(s)}{s}+\frac{(1-\widehat{\phi}(s))\widehat{h}_{\rm{eq}}(s)\widetilde{f}(k)}{s(1-\widehat{\phi}(s)\widetilde{f}(k))}
\end{split}
\end{equation}
This equation is a modification of the Montroll-Weiss equation, taking into consideration the
equilibrium initial state.
Using the Laplace transform of Eq. \eqref{19wesgf120}, we have
\begin{equation}\label{19sadeq101F1}
\begin{split}
  \widetilde{\widehat{P}}_{\rm{eq}}(k,s)=\frac{\langle\tau\rangle s-1+\widetilde{\phi}(s)}{\langle\tau\rangle s^2}+\frac{(1-\widehat{\phi}(s))^2 \widetilde{f}(k)}{\langle\tau\rangle s^2(1-\widehat{\phi}(s)\widetilde{f}(k))}.
\end{split}
\end{equation}
The first term on the right-hand side is
$k$ independent, hence its inverse Fourier transform gives a delta function on the initial condition $x=0$ describing non-moving particles.  This population of motionless particles is non negligible in the sense that they contribute to the MSD; see Eq.~\eqref{19sadeq103a1f}.
%
%

\begin{figure}[h]
 \centering
 \includegraphics[width=8cm, height=6.5cm]{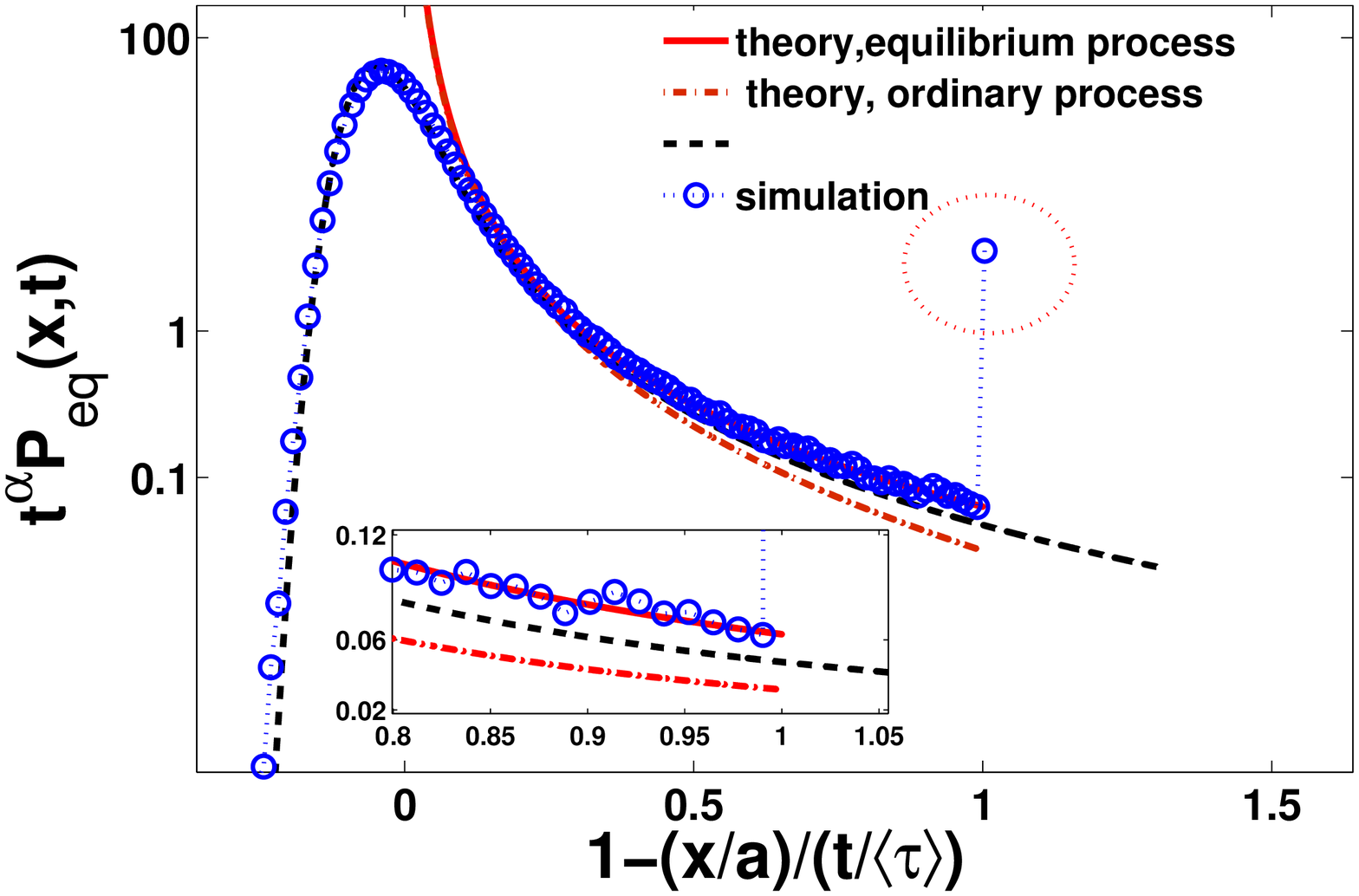}\\
 \caption{Scaled PDF of the position versus $1-(x/a)/(t/\langle\tau\rangle)$.
The symbols are simulation results obtained from $3\times10^6$ realizations. The red solid line calculated from Eq.~\eqref{19sadeq102fg} describing the behavior when $x\propto t$ is consistent with the far tail of simulation results; see  inset.
We also plot the  theory  of an ordinary process Eq.~\eqref{19seq13}, showing that it  clearly fails, and it under estimates the rare fluctuations described by the equilibrium  theory.
Notice that the delta like contribution circled in red, describes non-moving particles at $x=0$.
Here $\alpha=1.5$, $a=1$, $\sigma=1$, $t_a=10^4$, $t=1000$, and $\tau_0=0.1$.
%
%
%
%
}\label{EqrareEventsLongTime}
\end{figure}

Based on Eq.~\eqref{19sadeq101F1}, we consider  typical fluctuations, i.e., $k,s\to 0$ and $|k|\propto |s^\alpha|$
\begin{equation}\label{19sadeq101a1}
\begin{split}
\widetilde{\widehat{P}}_{{\rm eq }}(k,s) &\sim \frac{(1-\widehat{\phi}(s))^2}{\langle\tau\rangle s^2}\frac{1}{1-\widehat{\phi}(s)\widetilde{f}(k)} \\
    & \sim\frac{\langle\tau\rangle}{\langle\tau\rangle s-ika-b_\alpha s^\alpha},
\end{split}
\end{equation}
where we used the asymptotic behaviors of $\widetilde{\phi}(s)$ and $\widehat{f}(k)$.
The inverse Laplace-Fourier transform of   Eq.~\eqref{19sadeq101a1} yields
\begin{equation}\label{19sadeq101a2}
P_{\rm{eq}}(x,t)\sim\frac{1}{a(t/\overline{t})^{1/\alpha}}L_{\alpha,1}\left(\frac{x-at/\langle\tau\rangle}{a(t/\overline{t})^{1/\alpha}}\right),
\end{equation}
According to Eq.~\eqref{19sadeq101a2}, the typical fluctuations are the same as the one of the ordinary case; see Eq.~\eqref{19seq7} and the dashed lines in Fig.~\ref{EqrareEventsLongTime}. That is, the bulk fluctuations  do not depend on the initial state.
On the other hand, the MSDs of both cases are different, this means that the far tail of $P_{\rm{eq}}(x,t)$ should be modified compared with the ordinary case.
As mentioned before, the normalized density Eq.~\eqref{19sadeq101a2} gives an unphysical infinite MSD due to the slowly decaying tail of asymmetric L{\'e}vy distribution. This means that we expect modifications of this limiting law at the far tail.

For the rare events of the equilibrium CTRW, i.e., both $s$ and $k$ are small and comparable, inserting $\widehat{\phi}(s)$ and $\widetilde{f}(k)$ into Eq.~\eqref{19sadeq101F1} gives
\begin{equation}\label{19sadeq102}
\begin{split}
\widetilde{\widehat{P}}_{\rm{eq}}(k,s) &\sim \frac{b_\alpha}{\langle \tau\rangle s^{2-\alpha}}+\frac{\langle\tau\rangle -2b_\alpha s^{\alpha-1}}{\langle\tau\rangle s-ika-b_\alpha s^\alpha}.
\end{split}
\end{equation}
Rewriting the second term of the right-hand side of Eq.~\eqref{19sadeq102} as
\begin{equation}\label{19sadeq103tyr}
\frac{\langle\tau\rangle -2b_\alpha s^{\alpha-1}}{\langle\tau\rangle s-ika-b_\alpha s^\alpha}\sim \frac{\langle\tau\rangle-b_\alpha s^{\alpha-1}}{\langle\tau\rangle s-ika}+\frac{b_\alpha s^{\alpha-1}ika}{(\langle\tau\rangle s-ika)^2}
\end{equation}
and using the relation
\begin{equation}\label{19sadeq104}
\mathcal{F}^{-1}\mathcal{L}^{-1}\left[\frac{ s^{\alpha-1}}{\langle\tau\rangle s-ika}\right]=\frac{\left(t-\frac{\langle\tau\rangle}{a}x\right)^{-\alpha}}{a\Gamma(1-\alpha)},
\end{equation}
we get the main results of this section describing the packet when $x$ is of the order of $t$
\begin{equation}\label{19sadeq102fg}
P_{\rm{eq}}(x,t)\sim \frac{(\tau_0)^\alpha t^{1-\alpha}}{\langle\tau\rangle(\alpha-1)}\delta(x)+\frac{(\tau_0)^\alpha}{at^\alpha}\mathcal{I}_{{\rm eq},\alpha}(\xi),
\end{equation}
where the non-normalised state function reads
\begin{equation}\label{19sadeq102gr1}
\mathcal{I}_{{\rm eq},\alpha}(\xi)=\alpha \xi^{-\alpha-1}+(2-\alpha)\xi^{-\alpha}
\end{equation}
and $\xi=1-(x/a)/(t/\langle\tau\rangle)$. Comparing with Eq. \eqref{19seq13}, we see that the infinite densities for the equilibrium and non-equilibrium processes are  different. This indicates that initial conditions influence the statistics at small position even when the measurement time is long $t\gg \langle\tau\rangle$. 
The rare  fluctuations for the equilibrium case are larger if compared with the ordinary  process,
in particular they include a delta function contribution; see  the data marked in a red circle in Fig.~\ref{EqrareEventsLongTime}. This means that particles not moving at all  contribute to the rare events. Note that Eq.~\eqref{19sadeq102fg} can be matched to the far tail of the L\'{e}vy distribution Eq.~\eqref{19sadeq101a2}, as it should.

We further  check that the MSD is determined by the rare fluctuations Eq.~\eqref{19sadeq102fg} resulting in a different MSD compared with the ordinary process. Using the random variable $\eta=(x-at/\langle\tau\rangle)/(at/\langle\tau\rangle)$ with $-1<\eta<0$,
from Eq.~\eqref{19sadeq102fg} we get
\begin{equation}\label{19sadeq102c1}
\langle \eta^2\rangle_{{\rm eq}} \sim \frac{2b_\alpha t^{1-\alpha}}{\langle\tau\rangle\Gamma(4-\alpha)};
\end{equation}
see Appendix \ref{infinieyes}.
Similarly, $\langle \eta^2\rangle_{{\rm or}}$ is also obtained according to Eq.~\eqref{19seq13}.
Utilizing Eqs.~\eqref{19seq5} and \eqref{19sadeq102c1},
\begin{equation}\label{19sadeq102c2}
 \langle x^2\rangle-\langle x\rangle^2 \sim \displaystyle \left\{
                                                 \begin{array}{ll}
                                               \displaystyle \frac{2a^2b_\alpha(\alpha-1) t^{3-\alpha}}{\langle\tau\rangle^3 \Gamma(4-\alpha)} , & \hbox{ordinary;} \\
                                                 \displaystyle  \frac{2a^2b_\alpha t^{3-\alpha}}{\langle\tau\rangle^3 \Gamma(4-\alpha)}, & \hbox{equilibrium.}
                                                 \end{array}
                                               \right.
\end{equation}
Though the MSDs for both cases grow as power law $t^{3-\alpha}$, the MSD for the equilibrium case is  larger than the ordinary one. Since the mean of the first waiting time following Eq.~\eqref{19wesgf120} is infinite, the probability of  particles
experiencing a long trapping time increases rapidly compared with an ordinary situation. In turn, this considerably  yields inactive particles which are trapped
on the origin  for the whole observation time $t$ far lagging behind the mean.
Hence, the MSD for the equilibrium process has a deep relationship with the motionless particles; see Eqs.~\eqref{19sadeq103a1f}.
It is interesting to find that the MSDs for both cases are determined by the far tail of the packet described by the infinite densities. As expected, when $\alpha\to 2$, these two processes show normal diffusion  with no difference, so then the initial condition is unimportant.


\begin{figure}[h]
 \centering
 \includegraphics[width=8cm, height=6.5cm]{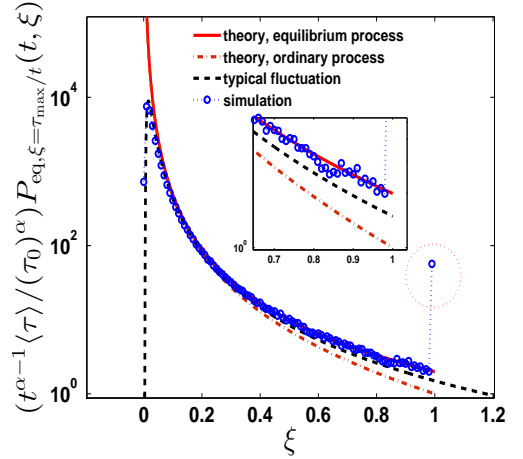}\\
 \caption{Simulations of the distribution of $\xi$ with the scaling $\xi=\tau_{\max}/t$ and $\alpha=1.5$ compared to the analytical predication obtained from Eq.~\eqref{19alwSec3qw1010}.  Clearly both the typical fluctuations Eq.~\eqref{19alwSec3qw107ewq} and the rare events Eq.~\eqref{19seq13} with an non-equilibrium condition do not work at the far tail of the distribution; see inset.
%
%
%
%
}\label{EqRareTam}
\end{figure}

\subsection{The rare fluctuations of $\tau_{\max}$}
After calculating $P(x,t)$ for small $x$, the next aim is to deal with the far tail of the PDF  $\tau_{\max}$ when $\tau_{\max}$ and $t$ are comparable.
From Eq.~\eqref{19alwSec3qw103}, we have
\begin{equation}\label{19alwSec3qw104}
\begin{split}
  \frac{1}{s}-\widehat{F}_{{\rm eq}}(s,L) & =\frac{1}{s(1+\widehat{G}(s,L))} \\
    & +\frac{1-\int_0^{L}\exp(-s\tau_1)h_{{\rm eq}}(\tau_1)d\tau_1}{s\left(1+\frac{1}{\widehat{G}(s,L)}\right)}\\
&-\int_0^{L}\exp(-sB_t)\int_{B_t}^\infty h_{{\rm eq}}(\tau)d\tau dB_t,
\end{split}
\end{equation}
where $\widehat{G}(s,L)$ is defined in Eq.~\eqref{19smeq1ggggg}.
It gives the PDF by the derivative
\begin{equation}\label{19alwSec3qw105}
 \widehat{P}_{{\rm eq},{\tau_{\max}}}(s,L)=-\frac{\partial [\frac{1}{s}-\widehat{F}_{\rm{eq}}(s,L)]}{\partial L}.
\end{equation}
Note that
\begin{equation}\label{19alwSec3qw106}
 \frac{\partial }{\partial L}\frac{1-\int_0^{L}\exp(-s\tau_1)h_{{\rm eq}}(\tau_1)d\tau_1}{s(1+1/\widehat{G}(s,L))}\sim \frac{\exp(-sL)}{-s}h_{{\rm eq}}(L)
\end{equation}
since $\widehat{G}(s,L)$ is large with $L\propto 1/s$. Using Eqs.~\eqref{19alwSec3qw105} and \eqref{19alwSec3qw106},  $\widehat{P}_{{\rm eq},\tau_{\max}}(s,L)$ reduces to
\begin{equation}\label{19alwSec3qw107}
\begin{split}
 \widehat{P}_{{\rm eq},{\tau_{\max}}}(s,L) &\sim \frac{1}{\widehat{G}(s,L)}+\frac{\alpha}{sL\widehat{G}(s,L)}\\
    &~~~ +\exp(-sL) \int_L^\infty h_{{\rm eq}}(\tau)d\tau\\
&~~~+\frac{\exp(-sL)}{s}h_{{\rm eq}}(L).
\end{split}
\end{equation}
Note that Eq.~\eqref{19alwSec3qw107} is a uniform approximation in Laplace space which is effective for numerous $L$ and large $t$. More exactly, within this approximation, we have the only condition that the observation time $t$ is large enough without considering the scaling between $t$ and $L$.
For the typical fluctuations,  the leading term of Eq.~\eqref{19alwSec3qw107} is the same as the ordinary process. Thus
 \begin{equation}\label{19alwSec3qw107ewq}
  \int_0^L P_{{\rm eq},{\tau_{\max}}}(t,y)dy
  \sim \exp\left[-\frac{t}{\langle\tau\rangle} \left(\frac{\tau_0}{L}\right)^\alpha\right].
 \end{equation}
We see that the typical fluctuations of the longest time interval of both the equilibrium and the ordinary renewal processes are the same and independent of the initial conditions, describing  the behavior when $L^\alpha$ is of the order of $t$.

Next we turn our attention to the case when $L\propto t$. Restart from Eq.~\eqref{19alwSec3qw107}, the inverse Laplace transform gives our main result describing the far tail of the density
\begin{equation}\label{19alwSec3qw108}
\begin{split}
  P_{{\rm eq},{\tau_{\max}}}(t,L) & \sim \frac{(\tau_0)^\alpha}{t^\alpha \langle\tau\rangle}\mathcal{I}_{{\rm eq},\alpha}(y)+\delta(t-L) \\
    & \times\int_{L}^\infty h_{{\rm eq}}(\tau)d\tau+\theta(t-L)h_{{\rm eq}}(L)
\end{split}
\end{equation}
with $L\leq t$.
Utilizing Eqs.~\eqref{friswairss} and \eqref{19alwSec3qw108}, we have
\begin{equation}\label{19alwSec3qw1010}
P_{{\rm eq},{\tau_{\max}}}(t,L)\sim\frac{(\tau_0)^\alpha}{t^\alpha \langle\tau\rangle}\mathcal{I}_{{\rm eq},\alpha}\left(\frac{L}{t}\right)+\delta(t-L)\frac{(\tau_0)^\alpha L^{1-\alpha} }{(\alpha-1)\langle\tau\rangle};
\end{equation}
see Fig.~\ref{EqRareTam}.
From Eqs.~\eqref{19sadeq102fg} and \eqref{19alwSec3qw1010}, it can be seen that the principle Eq.~\eqref{19seq2} is also valid for the equilibrium case. Though the typical fluctuations of $\tau_{\max}$ for equilibrium and ordinary process  show no difference, their far tails are distinct from each other [see Eqs.~\eqref{19seq4} and \eqref{19alwSec3qw1010}].

\section{Discussion and summary}

We have related the theory of extreme value statistics and the  fluctuations of a particle diffusing in a disordered system with traps. As mentioned, the observation of a non-Gaussian packet $P(x,t)$ and super-diffusive MSD is widely reported \cite{Shlesinger1974Asymptotic,Bouchaud1990Classical,Bouchaud1990Anomalous,Brian1997Anomalous,Berkowitz2006Modeling,Lechenault2010Super,Winter2012Active,Benichou2013Geometry,Schroer2013Anomalous,Gradenigo2016Field,Yong2016Backward,Leitmann2017Time,Pierre2018Nonequilibrium,Akimoto2018Ergodicity,Hou2018Biased}. Here we showed that a modification of Fr{\'e}chet's law is required to fully characterize these fluctuations. The largest waiting time  $\tau_{\max}$  is clearly shorter than the observation time $t$, namely the sum $\sum_{i=1}^N\tau_i+B_t$ is constrained, hence naturally we have deviations from the Fr{\'e}chet law. In other words, the theory of IID random variables, completely fails to describe the phenomenon of the far tail of the packet.  More profound is the observation that the statistics of $\tau_{\max}$ determines the far tail of $P(x,t)$ for the ordinary and equilibrium processes. One trapping event, the longest of the lot, controls the statistics of large deviations, and this is very different if compared  with standard large deviation theory \cite{Touchette2009large}, where many small jumps in the same direction control the statistics.

Our work is related to the so called single big jump principle, which was originally formulated for $N$ IID random variables $\{\vartheta_1,\vartheta_2,\cdots, \vartheta_N\}$ \cite{Cistjakov1964theorem}. It states that   $\sum_{i=1}^N\vartheta_i\doteqdot\max\{\vartheta_1,\vartheta_2, \cdots, \vartheta_N\}$ when the distribution of $\vartheta_i$ is sub-exponential, and for large maximum. Note that in the CTRW model considered in this manuscript we do not have any large spatial jump, instead we have long sticking events where the particles do not move. More importantly, in our case the waiting times are constrained  by the total measurement time and hence correlated, and their number $N$ fluctuates. Hence the situation encountered  here is simply different (though related) to the original one. Thus one aspect of our work was to modify the principle as we did in Eq.~\eqref{19seq2} and then describe the rare events with new  Eqs.~\eqref{19seq4}, \eqref{19seq13},  \eqref{19sadeq102fg}, and \eqref{19alwSec3qw1010}. This allowed us to connect the big jump theory to infinite densities.  The solutions describing the far tails of the distributions of $x$ and $\tau_{\max}$ are non-normalizable, still with proper scaling  they are the limits of the perfectly normalised  probability densities. For example in Eq.~\eqref{19seq4}, we multiply the normalized density $P_{{\rm or},{\tau_{\max}}}(t,L)$ by $\langle\tau\rangle(t/\tau_0)^\alpha$ and then get the infinite density $\mathcal{I}_{{\rm or},\alpha}(L/t)$. The variance of $\tau_{\max}$ and the  super-diffusive MSD are
calculated with these non-normalised states, meaning that these quantifiers of the anomaly  are sensitive to rare events.

We showed that the initial condition is an important factor controlling the behavior of the far tail
of distribution of interest. We calculated these for the stationary
and ordinary renewal processes, showing that for the stationary process motionless particles
give an important contribution to the
description of the rare fluctuations and the MSD. On the one hand this implies that the far tails are non-universal in their shapes. This can therefore be used to characterize the nature of the underlying  process.
As for universality, this shows up in the principle of big jump Eq.~\eqref{19seq2}, as the relation between the trapping time and the position $x$, is independent of the underlying process.

We note that the surprising super-diffusion of a biased tracer in a crowded medium was also found based on a many body theory \cite{Benichou2013Geometry,Liang2015Sample,Pierre2018Nonequilibrium}, diffusion of contamination in disordered systems,  and for numerical simulations of glass forming systems \cite{Winter2012Active,Schroer2013Anomalous} where it is interesting to check the relation of the dynamics and the big jump principle. The investigation of the single big-jump principle in the context of other models of random walks in random environments  is of great interest. For example the biased quenched trap model, exhibits typical fluctuations which are the same as those found for the biased CTRW \cite{Bouchaud1990Anomalous,Bouchaud1990Classical,Aslangul1991Two,Stanislav2017From}. Will this repeat for the rare events is still unknown.
Recently the case of $N$ IID random variables constrained to have a given sum was investigated, and under certain conditions the Fr\'{e}chet law was found \cite{Evans2008Condensation,Godreche2019Condensation,Majumdar2019Extreme}. From the constraint it is clear that the far tail of the distribution of maximum cannot be modeled with the Fr\'{e}chet law since there is a cutoff at the far tail. It would be of interest to investigate the far tail of this model (there $N$ was fixed while here $N$ is random) and see if the non-normalized density is found here as well.

~~\\
~~\\
~~\\
\begin{acknowledgments}
The authors would like to thank the anonymous reviewers for their helpful and constructive comments. EB acknowledges the Israel Science Foundation's grant
1898/17.

\end{acknowledgments}
\appendix
\begin{appendices}

\begin{widetext}

\section{
Calculation of E\lowercase{q}.~\eqref{19SEQ8}}\label{19smss2}
We now present the detailed derivation of Eq.~\eqref{19SEQ8} in the main text starting from the Montroll-Weiss Equation~\eqref{19seq5}.
Here we are interested in the case $x-at/\langle\tau\rangle\propto at/\langle\tau\rangle$ instead of $x-at/\langle\tau\rangle\propto at^{1/\alpha}$ describing the typical fluctuations (see the main text). In Fourier-Laplace space, this corresponds to $|s|\propto|k|$.
Plugging Eqs.~\eqref{18eq105} and \eqref{phits} into Eq.~\eqref{19SEQ8} leads  to
\begin{equation}\label{19smss21}
\begin{split}
  \widetilde{\widehat{P}}_{\rm or}(k,s) & \sim \frac{1}{\langle\tau\rangle s-ika-(\tau_0)^\alpha|\Gamma(1-\alpha)|s^\alpha}(\langle\tau\rangle-(\tau_0)^\alpha|\Gamma(1-\alpha)|s^{\alpha-1}) \\
    & =\frac{1}{(\langle\tau\rangle s-ika)(1-\frac{(\tau_0)^\alpha|\Gamma(1-\alpha)|s^\alpha}{\langle\tau\rangle s-ika})}(\langle\tau\rangle-(\tau_0)^\alpha|\Gamma(1-\alpha)|s^{\alpha-1}),
\end{split}
\end{equation}
where we use that $\langle\tau\rangle s$ and $ika$ are comparable, and neglect the term $k^2$ since $k^2\ll |s|$, $|k|$. Using $1/(1-y)\simeq 1+y$ with $y\to 0$, and  $s^{2\alpha-1}/(\langle\tau\rangle s-ika)\propto s^{2\alpha-2}\ll s^{\alpha-1}$, Eq.~\eqref{19smss21} reduces to
\begin{equation}\label{19smss22}
\widetilde{\widehat{ P}}_{\rm or}(k,s)\sim \frac{1}{\langle\tau\rangle s-ika}
  \left(\langle\tau\rangle-b_\alpha s^{\alpha-1}+\frac{(\tau_0)^{\alpha}|\Gamma(1-\alpha)|\langle\tau\rangle}{\langle\tau\rangle s-ika}s^\alpha\right).
\end{equation}
Regrouping, we have
\begin{equation}\label{19smss23}
\widetilde{\widehat{ P}}_{\rm or}(k,s)
 \sim \frac{1}{\langle\tau\rangle s-ika}
  \left(\langle\tau\rangle+\frac{iks^{\alpha-1}a(\tau_0)^{\alpha}|\Gamma(1-\alpha)|}{\langle\tau\rangle s-ika}\right),
\end{equation}
which gives  Eq.~\eqref{19SEQ8} in the main text.
\end{widetext}

\section{Moments of the position for equilibrium case}
We further consider the moments of the position for an equilibrium situation by using \cite{Klafter2011First}
\begin{equation}\label{19sadeq103a1}
\langle \widehat{x}^q(s)\rangle=(-i)^q\frac{\partial^q \widetilde{\widehat{P}}_{\rm{eq }}(k,s)}{\partial k^q}\Big|_{k=0}
\end{equation}
to check our theoretical result Eq.~\eqref{19sadeq102c1}.
For $q=1$, using Eqs.~\eqref{19sadeq101F1} and \eqref{19sadeq103a1}, we have
\begin{equation}\label{19sadeq103a2uy}
\langle \widehat{x}(s)\rangle_{{\rm eq}}=\frac{a}{\langle\tau\rangle s^2},
\end{equation}
from which yields
\begin{equation}\label{19sadeq103a3}
\langle x(t)\rangle_{{\rm eq}}=a\frac{t}{\langle\tau\rangle}.
\end{equation}
This is the exact result growing linearly with time $t$; see also Eq.~\eqref{19wesgf120a}.
Note that for an ordinary process, the asymptotic behavior of $\langle x(t)\rangle$ is $at/\langle\tau\rangle$.
When $q=2$, from Eq.~\eqref{19sadeq103a1} the second moment of $x(t)$ is
\begin{equation}\label{19sadeq103a4}
\langle x^2(t)\rangle_{{\rm eq}}\sim \frac{a^2 t^2}{\langle\tau\rangle^2}+\frac{2a^2b_\alpha t^{3-\alpha}}{\langle\tau\rangle^3\Gamma(4-\alpha)}.
\end{equation}
Utilizing Eqs.~\eqref{19sadeq103a3} and \eqref{19sadeq103a4}, the MSD is
\begin{equation}\label{19sadeq103a2}
\begin{split}
 \langle (x(t)-\langle x(t)\rangle)^2    \rangle_{{\rm eq}} & =\langle x^2(t)\rangle_{{\rm eq}}-\langle x(t)\rangle^2_{{\rm eq}} \\
    &\sim\frac{2a^2b_\alpha t^{3-\alpha}}{\langle\tau\rangle^3 \Gamma(4-\alpha)}.
\end{split}
\end{equation}
It gives that the  process shows super-diffusion, increasing  faster than the ordinary process. As expected, Eq.~\eqref{19sadeq103a2} is consistent with  Eq.~\eqref{19sadeq102c2} obtained from the infinite density Eq.~\eqref{19sadeq102fg}.

We further consider how  motionless particles contribute to the MSD. Taking the inverse Laplace-Fourier transform on the first term on the right-hand side of Eq.~\eqref{19sadeq101F1addf} gives $\int_t^\infty h_{{\rm eq}}(\tau)d\tau\delta(x)$. From Eq.~\eqref{friswairss} one can show that
\begin{equation}\label{19sadeq103a1f}
\begin{split}
&\langle (x-\langle x\rangle_{{\rm eq}})^2\rangle_{\rm{eq}}\\
&\geq \int_{-\infty}^\infty (x-\langle x\rangle_{\rm eq} )^2 \int_t^\infty h_{{\rm eq}}(\tau)d\tau \delta(x)dx\\
   &=\frac{a^2(\tau_0)^\alpha t^{3-\alpha}}{\langle\tau\rangle^3 (\alpha-1)},
\end{split}
\end{equation}
where we used the relation Eq.~\eqref{19sadeq103a3}. Since the MSD grows like $t^{3-\alpha}$, clearly this term describing non-moving particles controls the leading term of the MSD Eq.~\eqref{19sadeq103a2}.
While for the non-equilibrium case we get a
contribution of motionless particles to the MSD which increases like $t^{2-\alpha}$ and is negligible.

\section{The calculation of MSD\lowercase{s} using the infinite densities}\label{infinieyes}
In principle, the MSDs can  be calculated according to  Eq.~\eqref{19sadeq103a1}. However, in the long time limit it is easy to calculate MSDs based on the non-normalized density. This method is also valid  for high-order moments \cite{Wang2018Renewal}.
From Eq.~\eqref{19seq13} the scaling behavior of $\xi=1-(x/a)/(t/\langle\tau\rangle)$ gives
\begin{equation}\label{adddedss101}
  P_{\rm {or}}(\xi,t)\sim\frac{(\tau_0)^\alpha t^{1-\alpha}}{\langle\tau\rangle}\mathcal{I}_{\rm{or},\alpha}(\xi),
\end{equation}
where $0<\xi<1$. The second moment of $\xi$ is
\begin{equation}\label{adddedss102}
\begin{split}
  \langle\xi^2\rangle_{\rm{or}}& \sim \int_0^1\xi^2 P_{\rm {or}}(\xi,t) d\xi \\
    & =\frac{2(\tau_0)^\alpha t^{1-\alpha}}{\langle\tau\rangle(2-\alpha)(3-\alpha)}.
\end{split}
\end{equation}
and $\langle x(t)\rangle\sim at/\langle\tau\rangle$.
Using the relation $\langle\xi^2\rangle_{\rm{or}}=\langle(1-\frac{x/a}{t/\langle\tau\rangle})^2\rangle_{\rm{or}}$, we have
\begin{equation}\label{adddedss103}
\begin{split}
  \left\langle \left(x-\frac{at}{\langle\tau\rangle}\right)^2\right\rangle_{\rm{or}} & =\left\langle\left(a\xi\frac{t}{\langle\tau\rangle}\right)^2\right\rangle_{\rm{or}} \\
    & =\left(a\frac{t}{\langle\tau\rangle}\right)^2\langle(\xi)^2\rangle_{\rm{or}}\\
    &\sim \frac{2a^2(\tau_0)^\alpha t^{3-\alpha}}{\langle\tau\rangle^3(2-\alpha)(3-\alpha)}\\
    &=\frac{2a^2b_\alpha(\alpha-1) t^{3-\alpha}}{\langle\tau\rangle^3 \Gamma(4-\alpha)}.
\end{split}
\end{equation}
This means that the MSD of the non-equilibrium case is determined by the far tail of the density, i.e., the infinite density. Similarly, the MSD of an equilibrium process follows
\begin{equation*}\label{adddedss104}
\begin{split}
   \left\langle \left(x-\frac{at}{\langle\tau\rangle}\right)^2\right\rangle_{\rm{eq}} &
 \sim \int_{-\infty}^{\infty}\left(x-\frac{at}{\langle\tau\rangle}\right)^2\frac{(\tau_0)^\alpha t^{1-\alpha}}{\langle\tau\rangle(\alpha-1)}\delta(x)dx\\
    &~~~~+\left(a\frac{t}{\langle\tau\rangle}\right)^2\langle\xi^2\rangle_{\rm eq}\\
    &\sim\frac{2a^2b_\alpha t^{3-\alpha}}{\langle\tau\rangle^3 \Gamma(4-\alpha)}
\end{split}
\end{equation*}
with $\langle\xi^2\rangle_{\rm eq}=\int_0^1\xi^2 \frac{(\tau_0)^\alpha}{t^{\alpha-1}\langle\tau\rangle}\mathcal{I}_{{\rm eq},\alpha}(\xi)d\xi$.
Here we want to stress that in the case of integrable observables, one can use the non-normalized state described by the infinite density.
\end{appendices}


\end{document}